\documentclass[aps,pra,letterpaper,twocolumn,floatfix]{revtex4-2}
\usepackage{amsmath,amssymb} 
\usepackage[T1]{fontenc}
\usepackage[utf8]{inputenc} 

\usepackage{graphicx} 

\usepackage{hyperref}
\hypersetup{colorlinks=true,citecolor={black},linkcolor={blue},urlcolor={blue}} 
\usepackage{xcolor}
\usepackage{bm}

\newcommand{\apd}[1]{\hat{a}^{\vphantom{\dag}}_{+}}
\newcommand{\ve}{\varepsilon}

\newcommand{\normord}[1]{\vcentcolon\mathrel{#1}\vcentcolon}
\providecommand{\vcentcolon}{\mathrel{\mathop{:}}}

\graphicspath{{.}{./figs/}}

\begin{document}

\title{Quantum theory of polariton weak lasing and polarization bifurcations}

\author{Huawen Xu} 
\email{huawen.xu@ntu.edu.sg}
\affiliation{Division of Physics and Applied Physics, School of Physical and Mathematical Sciences, Nanyang Technological University, 21 Nanyang Link, Singapore 637371, Singapore}

\author{Timothy C. H. Liew}
\email{timothyliew@ntu.edu.sg}
\affiliation{Division of Physics and Applied Physics, School of Physical and Mathematical Sciences, Nanyang Technological University, 21 Nanyang Link, Singapore 637371, Singapore}

\author{Yuri G. Rubo} 
\email{ygr@ier.unam.mx}
\affiliation{Instituto de Energías Renovables, Universidad Nacional Autónoma de México, Temixco, Morelos, 62580, Mexico} 

\begin{abstract}
The quantum theory of polariton condensation in a trapped state reveals a second-order phase transition evidenced by spontaneous polarization parity breaking in sub-spaces of fixed polariton occupation numbers.  
The emission spectra of a polariton condensate demonstrate the coexistence of a symmetry-conserving condensate state with linear polarization and two symmetry-broken elliptically polarized states in the vicinity of the threshold. As a result, an oscillating linearly polarized second-order coherence $g^{(2)}_{xx}(t)$, with $g^{(2)}_{xx}(t)<1$ over some time intervals, is obtained. Spontaneous symmetry breaking is reflected in the second-order cross correlator of circular polarizations.
A related build-up of elliptically-polarized weak lasing also results in non-monotonous dependence of the circular second-order coherence on excitation power and interaction strength.
\end{abstract}

\date{\today}

\maketitle

\section{Introduction\label{sec:intro}}
The experimental realization of optically trapped polariton condensates \cite{cristofolini13,askitopoulos13} marked an important technological advance by allowing better control of condensate arrangements, polarization, and their mutual coupling.   
Positioning polariton condensates away from excitation spots in the microcavity plane reduces substantially thermal noise and condensate decoherence, thus narrowing polariton lasing lines in the frequency spectrum. Combined with the recent progress in increasing polariton lifetimes \cite{mukherjee21,estrecho21} this makes networks of trapped condensates a promising platform for all-optical devices \cite{ballarini13,sanvitto16} and neuromorphic computing \cite{berloff17,ohadi17,ballarini20,opala23,kavokin22}. 

Polariton condensates can encode information through their polarization (spin) state, which exhibits a great amount of interesting properties \cite{martin01,lagoudakis02,baumberg08,levrat10}. While at high excitation level linearly polarized condensates are formed, the polarization state near the condensation threshold is more complex. It was discovered \cite{ohadi15,dreismann16} that a single trapped polariton condensate undergoes parity breaking bifurcation leading to random formation of two possible states of elliptically polarized condensates with opposite handedness. The polarization bifurcation near the threshold can be understood as a 
\emph{weak lasing effect}, where coupled single polariton modes may condense in more than one collective many-body state due to the interplay of polariton-polariton repulsion and a small difference in the lifetimes of different normal modes (equivalent to dissipative coupling) \cite{aleiner12}. 
The mean-field theory used so far to describe this phenomenon studies the polariton condensation in terms of nonlinear driven-dissipative equations, where different condensate states correspond to fixed points or limit cycles \cite{kim20,ruizsanchez20} (sometimes referred to as time crystals \cite{nalitov19}). This theory lacks the ability to properly address the possible coexistence of condensates near the threshold and their quantum properties, which are speculated to allow a quantum speed-up of polariton simulators~\cite{lagoudakis17}. Moreover, the validity of the mean-field description of weak lasing is limited by the assumption of weak polariton-polariton interaction. An essential property of weak lasing is that the formation of a condensate is stabilized by the interaction, not by the depletion of an incoherent feeding reservoir. As a result, the condensate occupation numbers are inversely proportional to the interaction constant. The latter is controlled by the condensate confinement, so that the requirement of weak interaction is a severe limitation. 

In this paper we discuss the properties of the trapped polariton condensate density matrix in the presence of the weak lasing effect. The system in this case obeys open-dissipative quantum dynamics described by the Lindblad master equation for the density matrix. We analyze properties of both the steady state solution and different two-time correlators,
since the polarization and the many-body correlations can provide information on the presence of quantum fluctuations in the system \cite{bleu21,scarpelli24}.
We also calculate the emission spectra from the condensate, and show how they indicate the symmetry breaking transition in the system and coexistence of condensates with different symmetries.

\section{Formalism\label{sec:forma}}
For a trapped condensate, we denote by $\hat{a}_x$ and $\hat{a}_y$ the annihilation operators for X and Y linearly polarized states, respectively, and assume different dissipation rates, $\Gamma+\gamma$ and $\Gamma-\gamma$, from these states. 
We also account for the linear polarization dichroism between these states, with frequency splitting $\ve$, so that the Y linear polarization possesses the highest frequency and the lowest dissipation rate. This corresponds to the experimental conditions of Refs. \cite{ohadi15,dreismann16}.
If the polariton harvest rate $W$ is polarization independent, the condensate density matrix $\hat{\rho}$ evolves according to the equation
\begin{multline}\label{LindRho}
	\frac{d\hat{\rho}}{dt}\equiv\mathcal{L}\hat{\rho}=-i[\hat{H},\hat{\rho}] 
	-\frac{W}{2}\big([\hat{a}_x,\hat{a}_x^\dag\hat{\rho}]+[\hat{a}_y,\hat{a}_y^\dag\hat{\rho}]+\mathrm{H.c.}\big) \\
	-\frac{(\Gamma+\gamma)}{2}\big([\hat{a}_x^\dag,\hat{a}_x\hat{\rho}]+\mathrm{H.c.}\big)
	-\frac{(\Gamma-\gamma)}{2}\big([\hat{a}_y^\dag,\hat{a}_y\hat{\rho}]+\mathrm{H.c.}\big).
\end{multline}
Here the coherent evolution of the polariton condensate is given by the Hamiltonian of the Bose-Hubbard dimer \cite{laussy06} 
\begin{align}\label{BHHam}
	\hat{H}&= -\frac{\ve}{2}(\hat{a}^\dagger_+ \hat{a}_- + \hat{a}_-^\dagger \hat{a}_+) \notag \\
	&+\frac{\alpha_1}{4}(\hat{a}^{\dag2}_+\hat{a}^{2}_++\hat{a}^{\dag2}_-\hat{a}^{2}_-) 	+\frac{\alpha_2}{2}\hat{a}^\dag_+\hat{a}^\dag_-\hat{a}_+\hat{a}_-,
\end{align}
where we use the operators $\hat{a}_{\pm}=(\hat{a}_x\mp{i}\hat{a}_y)/\sqrt{2}$ for the circular polarization components, and $\alpha_1$ and $\alpha_2$ are the interaction constants for polaritons with the same and the opposite circular polarizations. 
The natural units of this system can be defined by taking $\Gamma$ as the unit of frequency and $\Gamma^{-1}$ as the unit of time.

The choice of a polarization independent harvest rate, $W$, follows previous theoretical models~\cite{read09}, which were established by comparison to experiment~\cite{ohadi12}, where the polarization independent pump rate does not itself prefer any specific polarization upon condensation. 
We note that possible initial linear polarization of the incoherent pump is typically lost during the process of relaxation of polaritons into condensate, since polariton-phonon scattering quickly randomizes the linear polarization component~\cite{roumpos09}.
Nevertheless, other factors may influence the condensate polarization. 
We have considered a real frequency splitting $\ve$ between X and Y linearly polarized states, and a specific implementation could be based on elliptical micropillar cavities~\cite{klaas19}, where X and Y linearly polarized states possess different spatial wave functions. Although we do not model the spatial distribution of the wave functions explicitly, it is expected that X and Y polarized states would have different penetrations through the boundaries of such cavities, resulting in different dissipation rates.

In what follows, it is convenient to introduce the spin of the condensate, which is defined by the operators
\begin{subequations}\label{SpinComp}
\begin{align}
	\hat{s}_1=\frac{1}{2}({\hat{a}_x^\dag}\hat{a}_x-{\hat{a}_y^\dag}\hat{a}_y), \quad 
	\hat{s}_2=\frac{1}{2}({\hat{a}_y^\dag}\hat{a}_x+{\hat{a}_x^\dag}\hat{a}_y), \label{SpinComp-a} \\
	\hat{s}_3=\frac{i}{2}({\hat{a}_y^\dag}\hat{a}_x-{\hat{a}_x^\dag}\hat{a}_y), \quad 
	\hat{s}_0=\frac{1}{2}({\hat{a}_x^\dag}\hat{a}_x+{\hat{a}_y^\dag}\hat{a}_y). \label{SpinComp-b}
\end{align}
\end{subequations} 
With the aid of spin components, the Hamiltonian \eqref{BHHam} can be written as 
\begin{equation}\label{LMGHam}
	\hat{H}=\hat{H}_0+\hat{H}_s, \qquad \hat{H}_s=-\ve{\hat s_1}+\frac{\alpha}{2}{\hat s_3^2},
\end{equation}
where $\alpha=\alpha_1-\alpha_2$. The Hamiltonian $\hat{H}_s$ is recognized as the Lipkin-Meshkov-Glick model \cite{lipkin65}, while $\hat{H}_0=[\alpha_1\hat{s}_0(\hat{s}_0-1)+\alpha_2\hat{s}_0^2]/2$ depends only on the total spin operator $\hat{s}_0$, which defines the total number of polaritons in the condensate. The dynamical properties of the Hamiltonians $\hat{H}$ and $\hat{H}_s$ are the same. It is seen from Eq.\ \eqref{LindRho} that the density matrix possesses the block-diagonal structure: the $1\times1$ block for the empty condensate, the $2\times2$ block for one polariton (spin 1/2), the $3\times3$ block for two polaritons (spin 1), etc. Each block evolves independently of the others under the Hamiltonian term in \eqref{LindRho}, but the neighboring blocks are coupled by the drive-dissipation Lindblad terms in \eqref{LindRho}.

\begin{figure}[t]
\includegraphics[width=1\columnwidth]{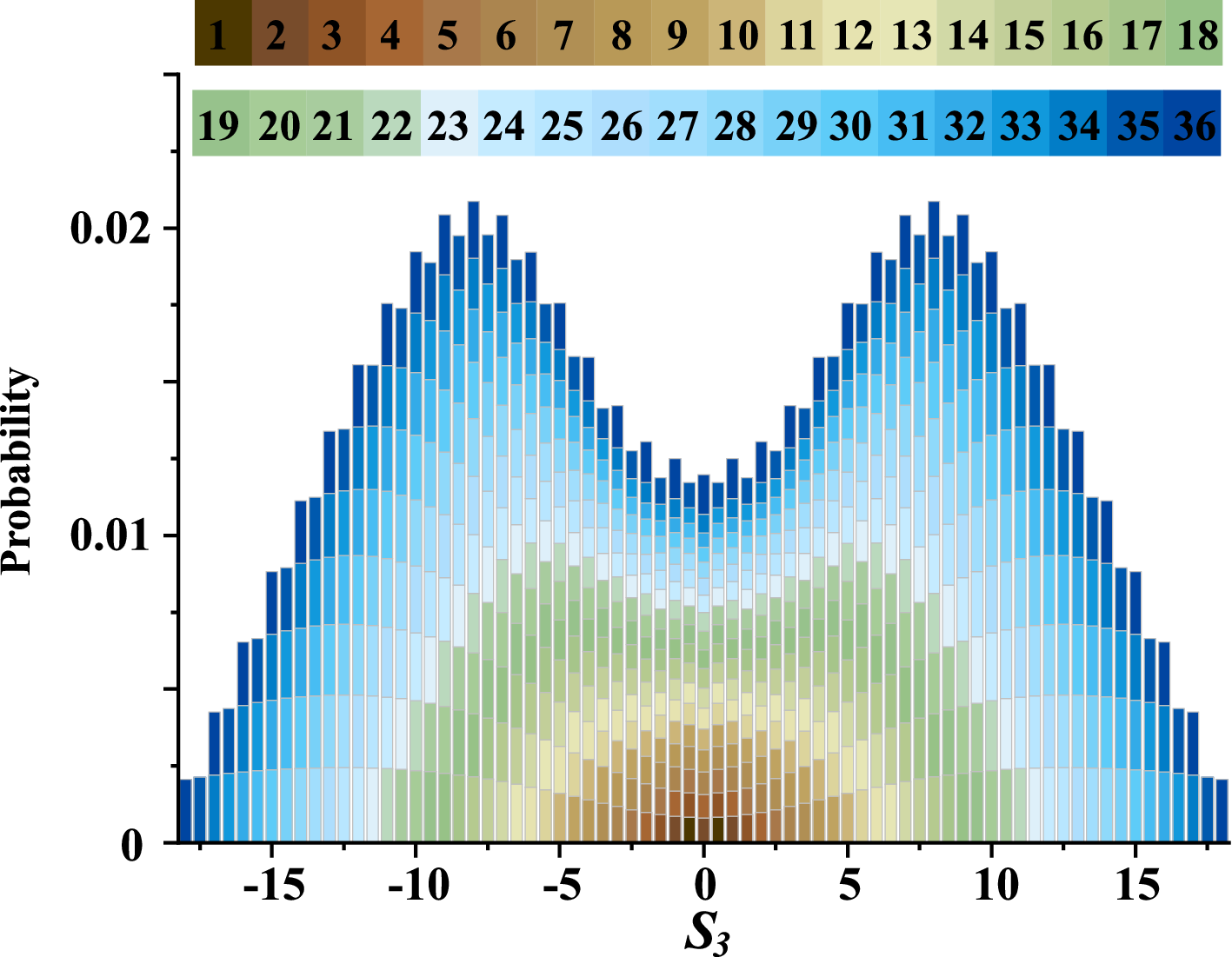}
\caption{Probability distribution of $s_3$ with different contribution from states with different number of particles, 
	where the numbers $1,\dots,36$ (encoded by different colors) indicate the number of particles 
	in a certain subspace of the system. The contribution from the vacuum state $|0,0\rangle$ is excluded. 
	We have considered a maximum of 36 particles on each site, $\gamma=0.5\Gamma$, $W=0.95\Gamma$, $\alpha=\ve=\Gamma$.}
\label{Fig:Histo}
\end{figure}

\section{Equilibrium condensate properties\label{sec:equil}}
One can note from Eq.~\ref{LindRho} that in this system, instead of having a specific condensation threshold, there is a threshold region from $W=\Gamma-\gamma$ to $W=\Gamma+\gamma$. Moreover, due to quantum fluctuations the condensate is already formed for $W<\Gamma-\gamma$. The semiclassical dynamics of the average spin components $S_\mu=\langle\hat{s}_\mu\rangle$ (see the Appendix for the details) reveals the presence of the second order phase transition. For small pumping $W$, the condensate is formed with the Y polarization, but for $W>W_c$ this state becomes unstable and two elliptically polarized condensates can appear with equal probability. This symmetry breaking manifests the weak lasing regime. The critical value $W_c$ is close to $\Gamma-\gamma$ for $\alpha\ll\Gamma$, and it shifts to smaller values with increasing interaction constant. 

The mean-field approximation describes the system well in the limit of large condensate populations, and for the weak lasing regime this is realized in the case of weak interaction, $\alpha\ll\Gamma$, see Eq.\ \eqref{WLScomp-d}. Here we will be interested in the case $\alpha\sim\Gamma$, where the exact quantum description is necessary. The regime of strongly interacting polariton condensates is not routinely achieved experimentally, but it has been discussed in the context of polariton blockade \cite{verger06}, it is a clear target in the field \cite{delteil19,munozmatutano19}, and it is expected in the near future, if not already achieved \cite{zhang21}. 

Clearly, a sharp phase transition only takes place in the limit of large condensate occupations, where the mean field approximation is valid. To investigate the polarization parity breaking phenomenon at strong interaction and low occupation numbers, we evolve the master equation \eqref{LindRho} until reaching the steady state $\hat{\rho}_0$. Then we check the probability distribution of the spin operator $\hat{s}_3$. In the circular $\hat{a}_{\pm}$ basis, it can be written as $\hat{s}_3=(\hat{a}^\dagger_{+}\hat{a}_{+}-\hat{a}^\dagger_{-}\hat{a}_{-})/2$ and it determines the polariton occupation number difference between the right and the left circularly polarized polariton modes. Possible eigenvalues of $\hat{s}_3$ are $s_3=0,\pm1/2,\pm1,\pm3/2,\dots$, and without parity breaking the $0$ eigenvalue has the highest probability (the polariton condensate is linearly polarized). 

The probability distribution $P(s_3)$ is given by
\begin{align}\label{ProbS3}
	P(s_3)&=\sum_{n_+,n_-}\langle n_+,n_-|\hat{\rho}_0|n_+,n_-\rangle \delta_{n_+,n_-+2s_3}  \notag \\ 
		  &=\sum_{n_+}\langle n_+,n_+-2s_3|\hat{\rho}_0|n_+,n_+-2s_3\rangle,
\end{align}
where $n_+$ and $n_-$ are the numbers of polaritons in the $+$ and $-$ states, respectively. We see from Fig.~\ref{Fig:Histo} that instead of centering at $0$, the probability distribution shows a double-peak structure, indicating that the polaritons have higher chance to condense at either the right circularly polarized ($s_3>0$) or the left circularly polarized ($s_3<0$) states. This is the fingerprint of the polariton condensate undergoing a circular polarization parity symmetry breaking.

In Fig.~\ref{Fig:Histo}, different colors in the stacked bar chart represent different contributions from states with different numbers of polaritons. For example, $1$ represents the contribution from states with only one polariton ($|0,1\rangle$ and $|1,0\rangle$). Also, one can note that states with more polaritons have a more pronounced symmetry breaking, as the interactions between the polaritons are more significant.

\begin{figure}[t]
\includegraphics[width=1\columnwidth]{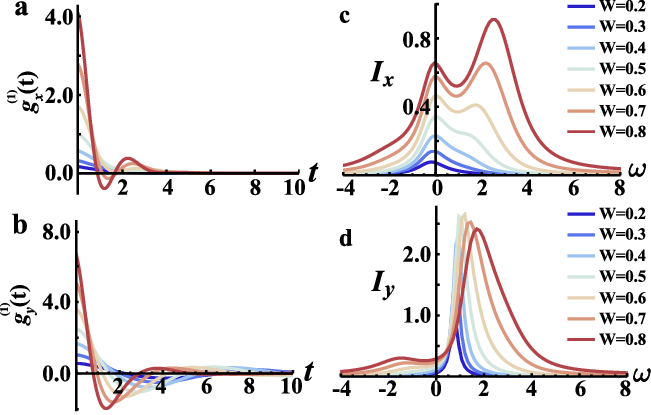}
\caption{Showing the first order correlation functions 
	for the X polarization (a) and the Y polarization (b), under different pumping strength $W$.  
	The corresponding emission spectra $I_{x,y}(\omega)$ are shown in panels (c) and (d). 
	We have considered a maximum of $14$ polaritons, $\gamma=0.5\Gamma$, $\alpha=\ve=\Gamma$. }
\label{Fig:Spectra}
\end{figure}

\begin{figure}[t]
\includegraphics[width=1\columnwidth]{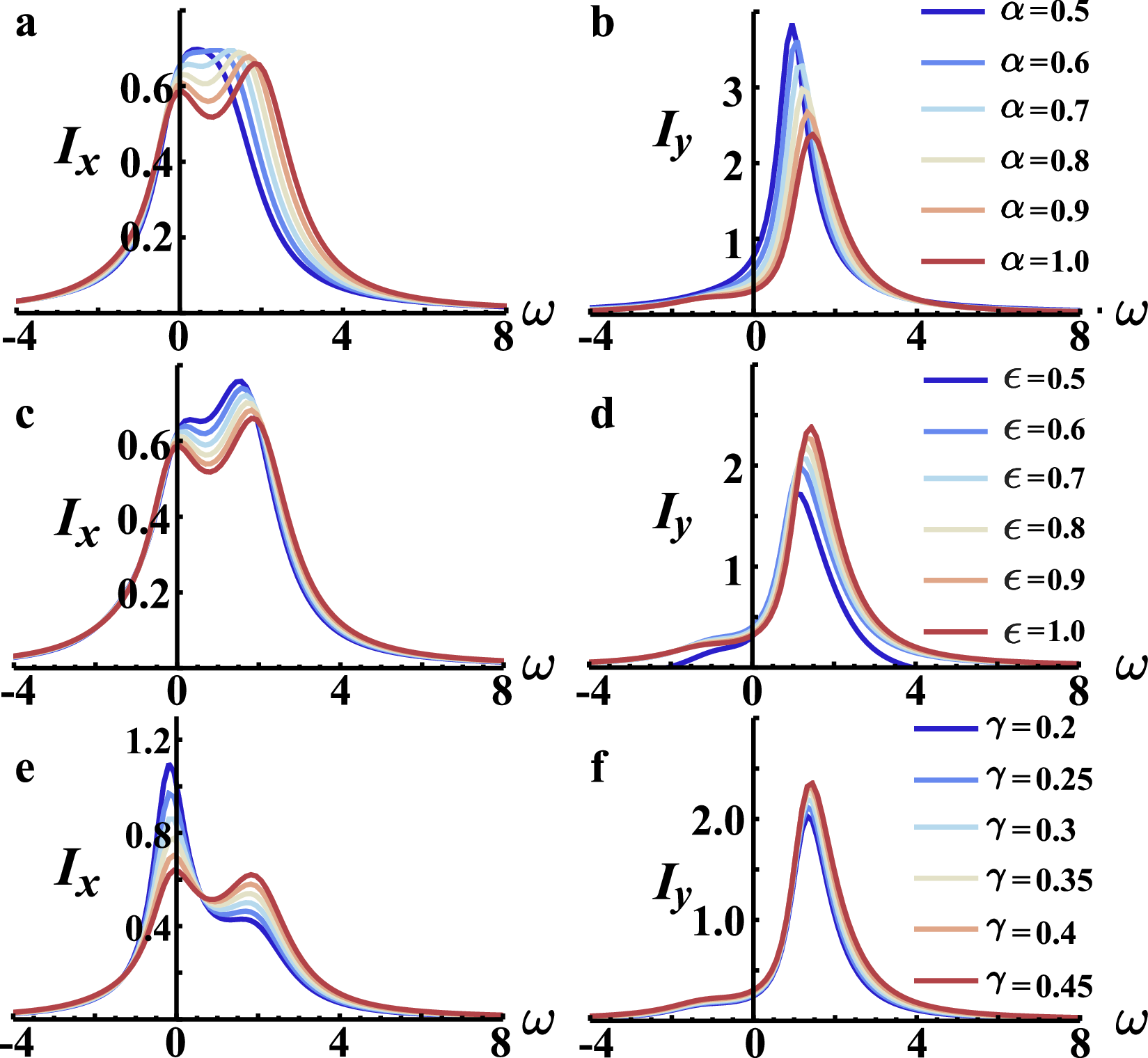}
\caption{Showing the dependences of the emission spectra in the X and Y linear polarization on the parameters of the condensate.
		For panels (a,b) $\gamma=0.5\Gamma$, $\ve=\Gamma$; for the panels (c,d), $\gamma=0.5\Gamma$, $\alpha=\Gamma$; 
		and for the panels (e,f), $\alpha=\ve=\Gamma$. 
		We have considered a maximum of 14 polaritons on each site and pumping strength $W=0.8\Gamma$ for all panels.}
\label{Fig:MoreSpec}
\end{figure}

\section{Emission spectra\label{sec:spectra}}
Here we analyze how the polarization symmetry-breaking transition is reflected in the light emission spectra, which could in principle be measured in an experiment. To this aim, we consider the time-delayed first-order correlation functions for X and Y polarized polariton modes,
\begin{equation}\label{G1oft}
	g^{(1)}_x(t)=\langle\hat{a}_x^\dag(t)\hat{a}_x(0)\rangle, \qquad
	g^{(1)}_y(t)=\langle\hat{a}_y^\dag(t)\hat{a}_y(0)\rangle.
\end{equation}
The time dependent correlators can be calculated in the framework of the quantum regression method \cite{lax63}.  Namely, to evaluate $\langle\hat{A}(t)\hat{B}(0)\rangle$ we evolve the modified operator $\hat{\rho}_B(0)=\hat{B}(0)\hat{\rho}_0$ to find $\hat{\rho}_B(t)=e^{\mathcal{L}t}\hat{\rho}_B(0)$ using the Lindblad super-operator $\mathcal{L}$ from Eq.\ \eqref{LindRho}. Then,  
$\langle\hat{A}(t)\hat{B}(0)\rangle=\mathrm{Tr}\{\hat{A}(0)\hat{\rho}_B(t)\}$. 

The first-order coherence functions \eqref{G1oft}, shown in Figs.~\ref{Fig:Spectra}(a,b), decrease with the time delay corresponding to a finite coherence time in the system. It is natural that they also oscillate in time given the energy difference of the X and Y modes, which is also renormalized by different blue-shifts of these modes with increasing pumping. It can also be noted that the first order coherence time decreases slightly with stronger pumping strength $W$. This is expected given that under higher pumping strength the interaction between the polaritons becomes more important and it leads to additional dephasing \cite{porras03}.


The emission spectra for X and Y polarizations can be found as
\begin{equation}\label{IofOm}
I_{x,y}(\omega)=\frac{1}{\pi}\mathrm{Re}\int_0^{\infty}g^{(1)}_{x,y}(t)e^{-i{\omega}t}dt,
\end{equation}
and they are shown in Figs.~\ref{Fig:Spectra}(c,d). These spectra demonstrate features that are not expected from mean-field theory presented in the Appendix. The latter predicts only one emission peak, originating either for the symmetry conserving Y-polarized state, or from two symmetry breaking elliptically polarized states, which possess opposite handedness but the same emission frequency. The X-polarization spectra, however, clearly demonstrate coexistence of two different condensate states and a corresponding two-peak structure. At low pumping, there is only one emission line originating from the X-polarized condensate, which appears initially at $\omega=-\ve/2$ and is gradually blue shifted with increasing pumping $W$. At moderate and strong pumping the second line becomes more and more important and it indicates the formation of weak lasing condensates. The blue shifts of different condensate states are different, with the blue shift of the weak lasing state being stronger, since it is more occupied. Consequently, at $W=0.8\Gamma$ we see the superposition of two peaks, separated by some distance larger than the Josephson splitting $\ve$. Similarly, in the spectrum of Y polarized polaritons (Fig.~\ref{Fig:Spectra}(d)), we can see also a weak second peak appearing at the low-frequency wing of the much stronger weak lasing peak. It is not quite resolved since the emission from the Y polarized polaritons is stronger, as they have smaller decay rate.

Additional properties of the emission spectra are shown in Fig.~\ref{Fig:MoreSpec}. The dependences of the emission intensity on the interaction strength, shown in Figs.~\ref{Fig:MoreSpec}(a,b), reveal the presence of the blue-shift with increasing interaction. The value of the blue-shift is bigger for the symmetry-broken condensate, so that the separation between the peaks in the X polarization increases as well.  
The two peaks acquire more visibility with increasing of the coherent $\ve$ and the dissipative coupling $\gamma$, as it is seen from Figs.~\ref{Fig:MoreSpec}(c,e).

\section{The second-order correlators\label{sec:secord}}
Another experimentally important quantity is the second-order coherence.  Here we calculate the time-dependent second-order correlation functions 
\begin{equation}\label{G2uv}
	g^{(2)}_{\mu\nu}(t)=\frac{\big<\hat{a}_\mu^\dag(0)\hat{a}_\nu^\dag(t)\hat{a}_\nu(t)\hat{a}_\mu(0)\big>}%
	{\big<\hat{a}_\mu^\dag(0)\hat{a}_\mu(0)\big>\big<\hat{a}_\nu^\dag(t)\hat{a}_\nu(t)\big>},
\end{equation}
where $\mu,\nu=x,y,+,-$.  
The numerator in Eq.\ \eqref{G2uv} can be again calculated using the quantum regression method, as the average $\langle\hat{A}(t)\hat{B}(0)\rangle=\mathrm{Tr}\{\hat{A}(0)\hat{\rho}_B(t)\}$ with $\hat{A}(0)=\hat{a}_\nu^\dag(0)\hat{a}_\nu(0)$, 
$\hat{\rho}_B(0)=\hat{a}_\mu(0)\rho_0\hat{a}_\mu^\dag(0)$, and $\hat{\rho}_B(t)=e^{\mathcal{L}t}\hat{\rho}_B(0)$. 

\begin{figure}[t]
\includegraphics[width=1\columnwidth]{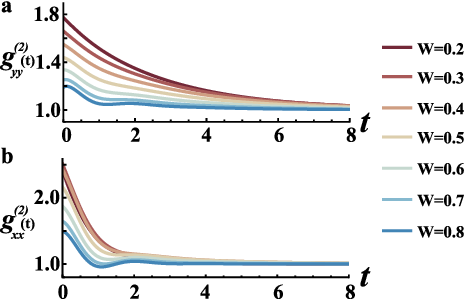}
\caption{The second order correlation function $g^{(2)}_{yy}(t)$ [Y polarization, panel (a)] 
	and $g^{(2)}_{xx}(t)$ [X polarization, panel (b)] for different pumping strengths $W$. 
	We have considered a maximum of 14 particles on each site, $\gamma=0.5\Gamma$, $\alpha=\ve=\Gamma$.}
\label{Fig:G2xy}
\end{figure}

The results for $g^{(2)}_{yy}(t)$ and $g^{(2)}_{xx}(t)$ are shown in Fig.~\ref{Fig:G2xy}. It can be seen that the second order coherence at zero time delay decreases from roughly $2$ (corresponding to a thermal state) toward $1$ (a coherent state) upon increasing the pumping strength. This is generally expected of polariton condensates, although we don't reach a perfect coherence due to the small numbers of particles involved. It can also be noted that the second order coherence of the vertical polarization (the $\hat{a}_y$ mode) exceeds that of the horizontal polarization (the $\hat{a}_x$ mode), and that there is an oscillation in the second order coherence taking place at higher pumping powers. Such oscillations are expected in coupled mode systems (e.g., in unconventional blockade systems~\cite{liew10}). Given that the mechanism in our present case of coupling $\hat{a}_x$ and $\hat{a}_y$ modes is via the nonlinear interaction term ($\alpha$), it is reasonable that these oscillations only become significant at higher pumping powers. Let us further recall that the $\hat{a}_y$ mode corresponds to the lower loss state and hence the state with higher number of particles. The $\hat{a}_y$ mode acts then as a source of particles for the $\hat{a}_x$ mode (even though the $\hat{a}_x$ mode also receives particles directly from the pumping). We note that this source corresponds to terms of the form $\hat{a}_x^\dag\hat{a}_x^\dag\hat{a}_y\hat{a}_y$, which appear from the $\hat{s}_3^2$ term. As the scattering rate of such a source term depends on the number of particles already in the $\hat{a}_x$ state this term can be considered as a stimulated scattering term, which consequently imparts coherence to the mode being developed. This explains why the $\hat{a}_x$ mode tends to have a smaller second order correlation function. Further, we note that it was previously shown in systems of bosonic cascades~\cite{liew16} that when a mode acquires coherence from another, the original source mode, which in our case is $\hat{a}_y$, should show more bunching. That is, $\hat{a}_y$ can develop a higher second order correlation function as a result of the nonlinear coupling.
Another interesting feature seen in Fig.~\ref{Fig:G2xy}(b) is the presence of time intervals with $g^{(2)}_{xx}(t)<1$. This effect becomes more pronounced with increasing interaction, but it is still not sufficient to obtain the sub-Poissonian polariton statistics.

\begin{figure}[t]
\includegraphics[width=1\columnwidth]{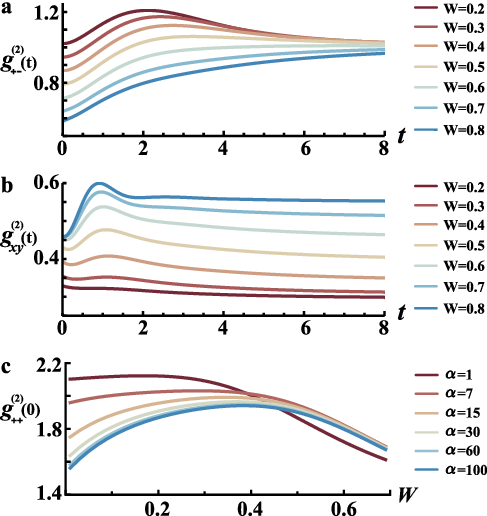}
\caption{(a) The time dependence of the circularly polarized second-order cross correlator $g^{(2)}_{+-}(t)$ 
	for different pumping strength $W$. 
    (b) The time dependence of the second-order cross correlator for linear polarizations $g^{(2)}_{xy}(t)$ 
	for different pumping strength $W$. 
	(c) The second-order coherence $g^{(2)}_{++}(0)$ for the same circular polarization.
	We have considered a maximum of 14 particles on each site, $\gamma=0.5\Gamma$, $\ve=\Gamma$ in all the panels, 
	and $\alpha=\Gamma$ in panels (a,b).}
\label{Fig:G2cc}
\end{figure}

The polarization parity symmetry breaking can also be evidenced from the second order cross correlator $g^{(2)}_{+-}(t)=g^{(2)}_{-+}(t)$ between the opposite circular polarizations, providing information about temporal correlations between the emission of photons with opposite handedness from the condensate. This correlator 
is shown in Fig.\ \ref{Fig:G2cc}(a). The value $g^{(2)}_{+-}(t)<1$ tells us that if a photon is observed in one circular polarization, then it is less likely to detect another photon in the opposite circular polarization. For long time delay, this anti-correlation is washed out, as the right and the left circular polarization have equal realization probabilities. 

Interestingly, similar anti-correlation is built-up with increasing pumping in the cross-correlator $g^{(2)}_{xy}(t)<1$ between X and Y linear polarizations, shown in Fig.\ \ref{Fig:G2cc}(b). In contrast, this cross-correlator does not saturate at 1. Its saturation value is close to that defined by the elliptical polarization of the symmetry-broken condensate state. We note that $g^{(2)}_{xy}(t)$ shows a peak at about the same time when $g^{(2)}_{xx}(t)$ shows a dip at corresponding pumping strength.

The results on the time dependence of the second-order cross correlators show that in our system the switching between different condensate states in the vicinity of the threshold takes place on a rather short time scale, about a few $\Gamma^{-1}$. This is because we considered strongly localized, trapped condensates, where the spatial degrees of freedom are ``frozen out'', while the quantum polarization fluctuations are strong. Recent state-of-art experiments \cite{alnatah24} have accessed fluctuations in spatially resolved condensates trapped in a wide disk and demonstrated formation of long-living space-polarization patterns in the vicinity of the threshold. It is interesting to extend the quantum theory of weak lasing to this case, but the analysis of such a problem is rather hard computationally.

In Fig.\ \ref{Fig:G2cc}(c) we also show the zero-delay second-order coherence for the same circular polarization $g^{(2)}_{++}(0)=g^{(2)}_{--}(0)$.
For small pumping, the second order coherence initially increases above 2 for weak interaction strength $\alpha$, but it becomes substantially smaller than 2 for strongly interacting polaritons. In this case, the dependence on the pumping $W$ is non-monotonous, with $g^{(2)}_{++}(0)$ approaching unity for large pumping, which manifests the polariton lasing.

\section{Conclusions\label{sec:concl}}
The theory of polariton condensation and spontaneous polarization formation is typically treated within the mean-field approximation, which explicitly assumes large occupation numbers of the condensates.  
In applications of polariton condensates it is often speculated that quantum fluctuations play a key role \cite{alnatah24}, and they can lead to quantum speedup of polaritonic devices~\citep{lagoudakis17}. Consequently, it is essential to generalize the fundamental theory of polariton condensation into the quantum realm accounting for such fluctuations.
By performing exact quantum calculations of the condensate density matrix, the first and the second order coherence, as well as the emission spectra, we demonstrate that the condensate formation near the threshold is indeed more complex. In particular, it involves possible coexistence of the states with broken and unbroken parity symmetry. The weak lasing in the quantum case is manifested by spontaneous polarization parity breaking across the different particle number subspaces, and it can be experimentally detected by measuring the second-order cross correlations of circular polarization components.

\section*{Acknowledgements}
H.X. and T.C.H.L. were supported by the Ministry of Education (Singapore) Tier 2 Grant MOE-T2EP50121-0020. 
This work was also supported in part by PAPIIT-UNAM Grant No.\ IN108524.

\appendix*
\section{Semiclassical approximation\label{sec:app}}
The Lindblad master equation for the density matrix \eqref{LindRho} can be used to obtain the equations for the dynamics of the averages of the spin operators $\hat{s}_\mu$, $\mu=0,1,2,3$, that are defined by Eqs.\ \eqref{SpinComp-a} and \eqref{SpinComp-b}. In particular, for the Hamiltonian \eqref{LMGHam}, that we use in this paper, the averages $S_\mu=\langle{\hat{s}_\mu}\rangle\equiv\mathrm{Tr}\{\hat{\rho}{\hat{s}_\mu}\}$ define the Stokes vector components and they satisfy the equations
\begin{subequations}\label{SVecDyn}
\begin{align}
	\dot{S}_0 &= -(\Gamma-W)S_0-{\gamma}S_1+W, \label{SVecDyn-a} \\
	\dot{S}_1 &= -(\Gamma-W)S_1-{\gamma}S_0-{\alpha}S_{23}, \label{SVecDyn-b} \\
	\dot{S}_2 &= -(\Gamma-W)S_2+{\ve}S_3+{\alpha}S_{13}, \label{SVecDyn-c} \\
	\dot{S}_3 &= -(\Gamma-W)S_3-{\ve}S_2, \label{SVecDyn-d}
\end{align}
\end{subequations}
where we also introduced the averages of the normal ordered products $S_{\mu\nu}=\langle\normord{\hat{s}_{\mu}\hat{s}_{\nu}}\rangle$. 

\begin{figure}[t]
	\centering 
	\includegraphics[width=0.99
    \columnwidth]{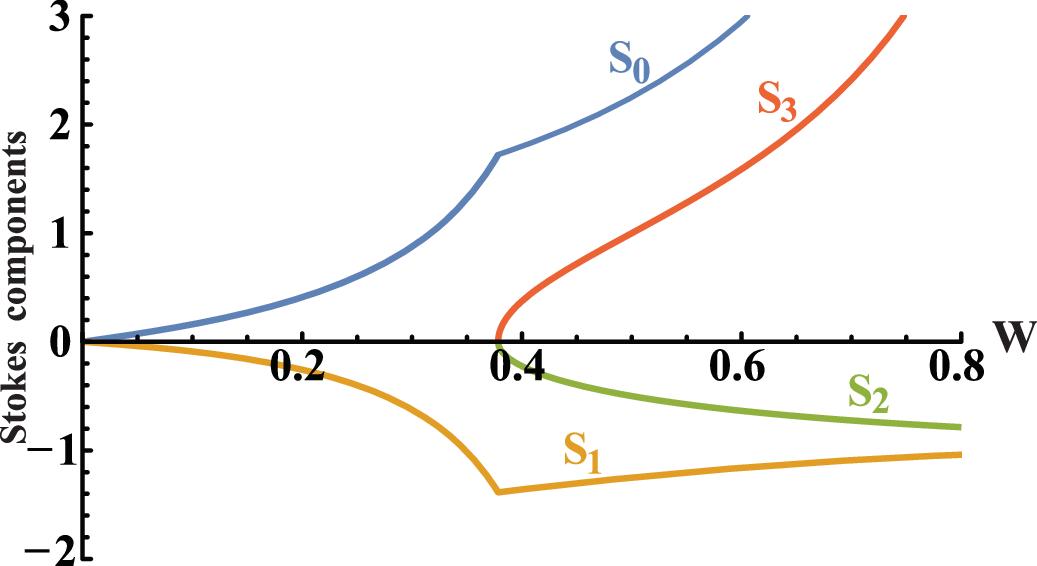} \caption{
	Showing the dependences of the spin components on the pumping $W$. 
	The parameters are $\Gamma=1$, $\gamma=0.5$, $\ve=\alpha=1$, 
	and they correspond to the bifurcation value $W_c\approx0.378$. }
    \label{Fig:PFork}
\end{figure}

The dynamics of the tensors $S_{\mu\nu}$ involves in turn the averages $S_{\mu\nu\lambda}=\langle\normord{\hat{s}_{\mu}\hat{s}_{\nu}\hat{s}_{\lambda}}\rangle$, and so on. The simplest way to break this chain of equations is by replacing $S_{23}\,{\rightarrow}\,S_2S_3$ and $S_{13}\,{\rightarrow}\,S_1S_3$ in Eqs.\ \eqref{SVecDyn-b} and \eqref{SVecDyn-c}, which explicitly assumes the second-order coherence to be equal to 1. Below we present the results of this approximation by analyzing the bifurcations of the fixed points of Eqs.\ \eqref{SVecDyn} under these substitutions in the region of pumping rates $0<W<\Gamma$, assuming also $0<\gamma<\Gamma$ and $\alpha\ve>0$.

For small $W$ there is only one static solution, which corresponds to formation of the condensate in the Y linearly-polarized state, defined by the $\hat{a}_y$ annihilation operator. The values of the spin components are
\begin{subequations}\label{YState}
\begin{align}
	S_0 &= \frac{W}{2}\left(\frac{1}{\Gamma+\gamma-W}+\frac{1}{\Gamma-\gamma-W}\right), \\
	S_1 &= \frac{W}{2}\left(\frac{1}{\Gamma+\gamma-W}-\frac{1}{\Gamma-\gamma-W}\right)<0, \\ 
	S_2 &= S_3 = 0.
\end{align}
\end{subequations}
It can be shown by making small perturbation of equations around this fixed point and performing linear stability analysis that this condensate becomes unstable for $W>W_c$, where critical pumping value $W_c$ corresponds to $\alpha\ve|S_1|=(\Gamma-W)^2+\ve^2$ and can be found as the root of equation
\begin{equation}\label{EqWc}
	[(\Gamma-W_c)^2-\gamma^2][(\Gamma-W_c)^2+\ve^2]=\alpha\gamma{\ve}W_c.
\end{equation} 

There is a pitchfork bifurcation at $W=W_c$, which results in the appearance of two stable symmetry breaking fixed points for $W>W_c$, describing the possibility of formation of two elliptically polarized condensates with positive ($S_3>0$) and negative ($S_3<0$) handedness. The components of the Stokes vector in this case are
\begin{subequations}\label{WLScomp}
\begin{align}
	S_1 &= -\frac{g^2+\ve^2}{\alpha\ve}, \label{WLScomp-a} \\ 
	S_2 &= -\frac{g}{\ve}S_3, \label{WLScomp-b} \\
	S_3 &= \pm\frac{1}{{\alpha}g}\sqrt{\alpha\gamma{\ve}W-(g^2-\gamma^2)(g^2+\ve^2)}\,, \label{WLScomp-c} \\
	S_0 &= \frac{1}{g}\left[W+\frac{\gamma(g^2+\ve^2)}{\alpha\ve}\right], \label{WLScomp-d}
\end{align}
\end{subequations}
where $g=\Gamma-W$.
The Stokes components as functions of $W$ are shown in Fig.\ \ref{Fig:PFork}, taking the positive value of $S_3$ in Eq.\ \eqref{WLScomp-c}.

%
%

\end{document}